# Pathways to Quantum Science for High-School and Incoming College Students

Dan-Adrian German✉
dgerman@iu.edu
Indiana University
Bloomington, Indiana, USA

Rebekah Randall
Canterbury High School
Ft. Wayne, Indiana, USA

Charles Pope
Indiana University Bloomington
Bloomington, Indiana

Michele Roberts
Indiana University
Bloomington, Indiana, USA

John Phillips
Indiana University
Bloomington, Indiana, USA

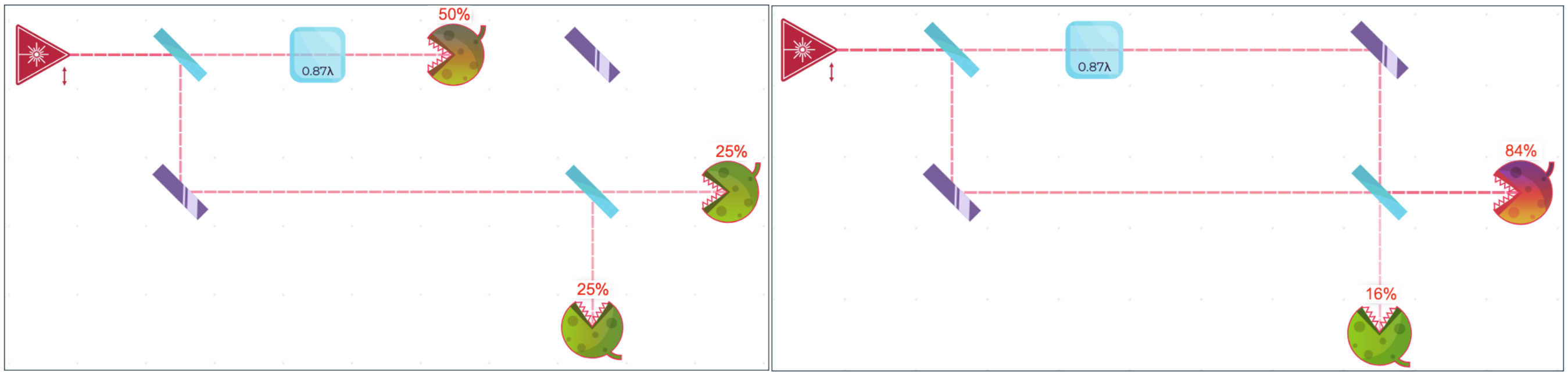

**Figure 1: Single-particle interference is a foundational concept in quantum computing.**

## Abstract

The convergence of AI and quantum computing requires a new approach to cybersecurity. Credible experts estimate that by 2030 a cryptographically relevant quantum computer will be capable of breaking the encryption that underpins all of our digital communication. One of the most disruptive technology in history is coming and it's expected to change everything. In this context academic institutions will contribute to their states' future economic prosperity by leveraging their strengths and resources, particularly talent development, in key and emerging areas prioritized across the state. We report on current efforts to develop a Pathways Plus to Quantum Information Science (QIS) degree, a set of dual credit courses for high-school (HS) students and incoming college students pursuing a minor or specialization in Quantum Computing (QC). As of today, in the US, quantum topics appear on the HS curriculum standards in just two states: OH (Computing) and TX (Physics). Reaching out to HS and even middle school students (and their teachers) presents obvious long-term benefits in terms of workforce development for the quantum industrial ecosystem. Standing in the way of successful, widespread introduction of QC and QIS topics in HS and middle school are three distinct obstacles: (a) lack of materials at the right level for students and instructors, (b) funding and support for professional development for teachers, and (c) lack of state standards. In this paper we address all three aspects with a special focus on the benefits of quantum virtual labs and ZX calculus in the classroom.

## 1 Introduction

Quantum computing has been front and center in the news for the better part of the last ten years. Recently, in an editorial in Wall Street Journal (WSJ) quantum computing is described as "a modern-day Manhattan Project, again a partnership between the U.K. and the U.S." Significant investments have been (and continue to be) made in China, France, Germany, Australia, Japan, the UK and, of course, the United States. In the U.S. the Quantum Economic Development Consortium (QED-C) is a premier, industry-driven organization (established in 2019) aimed at enabling and growing the global quantum technology industry and its supply chain. They have published a collection of three reports [14] in April on the state of the global quantum industry, a quantum computing and sensing markets forecast, as well as information on their methodology and an overview. The industry is maturing, the talent remains scarce.

*Submitted to ACM SIGCSE '27, Sacramento, CA*

And as the WSJ article puts it "[just a]s in 1943, most people haven't yet grasped what is coming. When they do, the countries that moved first will hold advantages and those that didn't will discover that the future belongs to those who build it themselves." [18].

## 2 The Challenge

Quantum physics is known to be challenging for two reasons: it describes counter-intuitive phenomena and employs rather advanced mathematics. Scarani [17] laments the apparent difficulty as follows: "Many notions of modern science have found their way into school programs all over the world. [...] But this is not the case for quantum physics, an unrivaled body of knowledge, both in the precision of its predictions and in the breadth of its scope. [...] In spite of all these, school programs cautiously keep away from quantum physics. Why is this so? There is an obstacle in the presentation of quantum physics, and a huge one at that: one cannot make an image of quantum phenomena." He then goes on to mention superposition, interference, entanglement as well as the wave-particle duality (all of them quintessential aspects of quantum theory) and concludes: "There is no image for any of these. [... So:] how can one learn quantum physics?" Well, we're here to report (16 years later) on the existence and success of new tools that greatly simplify, and alleviate this problem: the Quantum Flytrap (a virtual environment for quantum experiments online), the Misty States Formalism (developed and published by Terry Rudolph, in 2017) and Quantum in Pictures (or, the ZX calculus [10, 15]) developed over 20+ years at the University of Oxford (U.K.) all of which have been shown to be successful [1, 2, 5, 11] in communicating the unadulterated (and entirely operational) essence of quantum theory and quantum computation to young (and very young) audiences.

## 3 Misty States

It is important to note that our focus is (and should be) limited to quantum computing and not quantum mechanics in general, or quantum physics as a whole. Learning quantum computing is much easier [3] than learning quantum mechanics because quantum computing deals with a simple subset of quantum mechanics, as follows: (a) a qubit—the foundation of quantum computing—is the simplest non-trivial quantum system; (b) you never have to solve the Schrödinger equation or even learn what it is, because the quantum systems that carry out quantum computations evolve in a controlled manner based on the quantum gates applied to them; and (c) there's already a model of quantum computation, so the most difficult aspect of quantum mechanics—the art of applying it to real systems—is absent. Now consider the following well-formed misty state: {🍎, {🍉, 🍒}}. It says that three type of fruit are possible for a snack: apple, watermelon and cherries. The specific item that we end up with is determined probabilistically (via measurement). We do know it will be one of the three items listed. In this misty state the watermelon and cherries are in equal superposition with each other (let's call that state $s_1$) and the apple is in equal superposition with the state $s_1$. When we estimate the probability of each outcome we find that the probability of receiving an apple is $p(\text{🍎}) = \frac{1}{2}$ while for the other two items $p(\text{🍉}) = p(\text{🍒}) = \frac{1}{4}$ (so each has a probability of 0.25). If we modify the misty state to {🍎, {🍎, 🍒}} we have a minimal non-classical situation. One would expect the probability of getting an apple to be 0.75, when in effect it becomes 0.853 (thus lowering the chance of receiving cherries to 0.147). The reason, of course, is that the probability amplitudes add up, but the probabilities do not. The resulting state is:

$$|\Psi\rangle = \frac{\sqrt{2+\sqrt{2}}}{2}|\text{🍎}\rangle + \frac{\sqrt{2-\sqrt{2}}}{2}|\text{🍒}\rangle = \cos\frac{\pi}{8}|\text{🍎}\rangle + \sin\frac{\pi}{8}|\text{🍒}\rangle$$

One can demonstrate this experimentally in the virtual lab environment provided by the Quantum Flytrap, as shown in Fig. 1.

## 4 Quantum Flytrap

Let's now show how Quantum Flytrap can be used to implement the Elitzur-Vaidman experimental setup that allows the detection of a super-sensitive explosive device without setting it off.

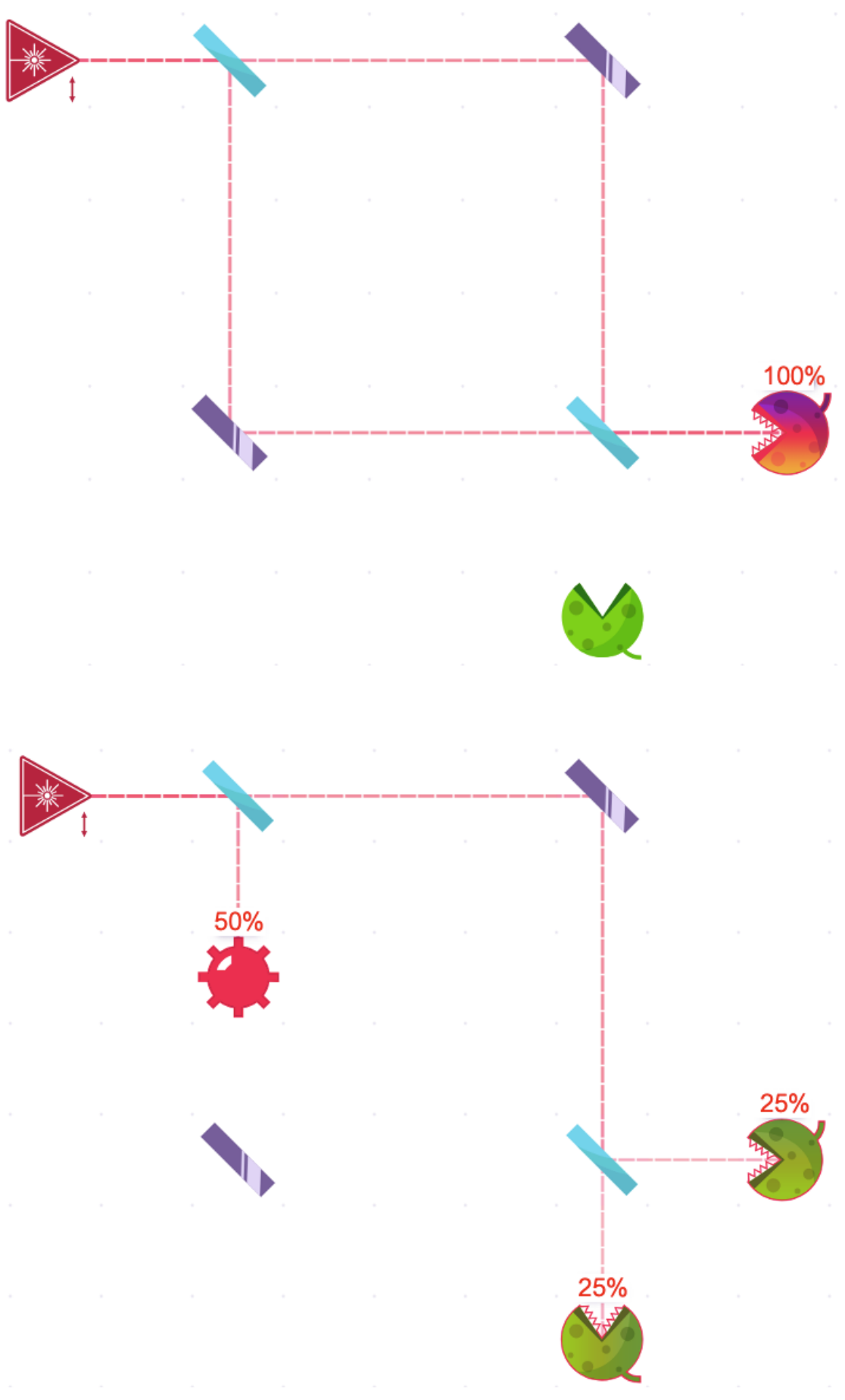

**Figure 2: Chance of detecting bomb without explosion: 25%.**

Here, as is also recommended in the CS2023 Knowkedge Unit [4], we only promote an appreciation for, and familiarity with, the main quantum concepts. But our approach is unabashedly hands-on. A classroom-tested text using misty states is available at [7]. It has been used with great success in the past at SIGCSE [8] in summer camps, QIS Faculty Learning Communities for HS and middle school teachers and now with dual-credit classes [5], [6].

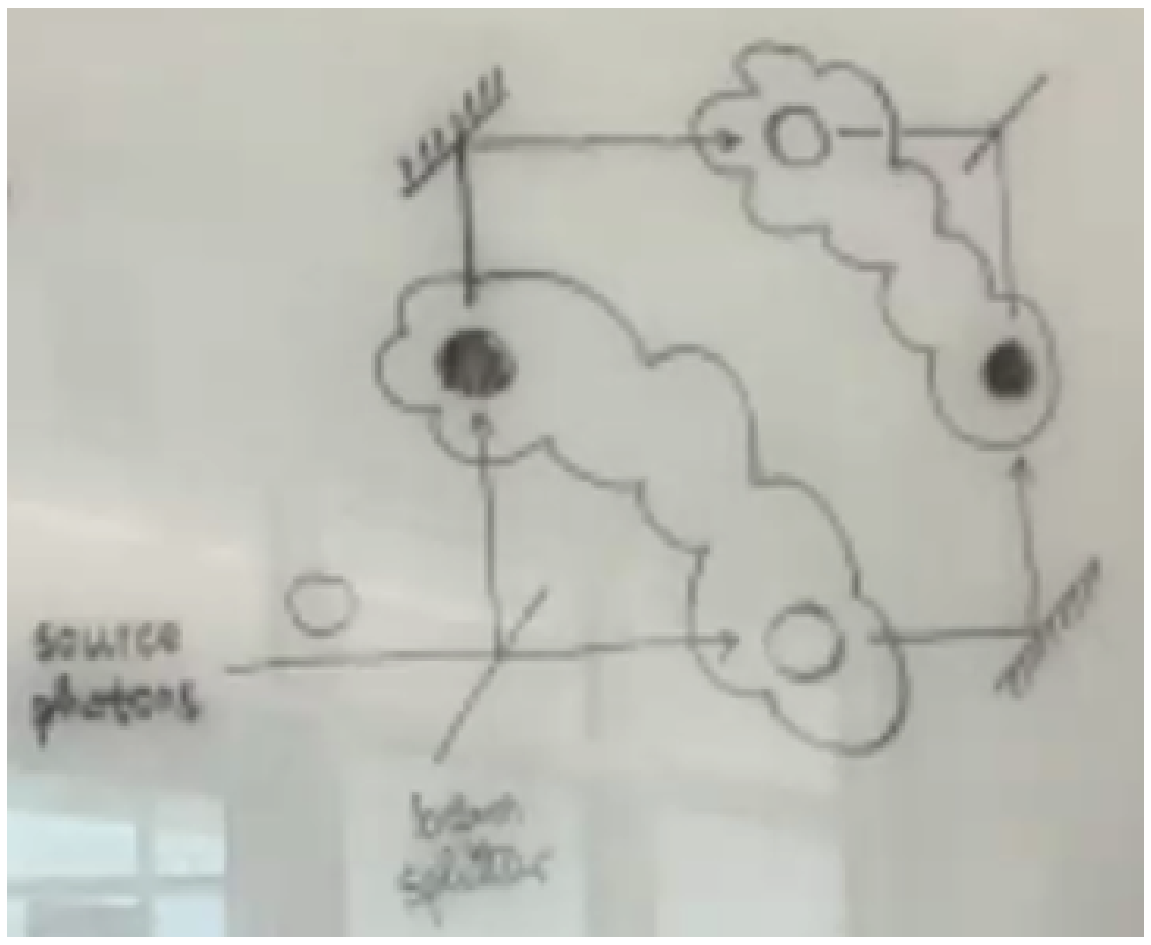

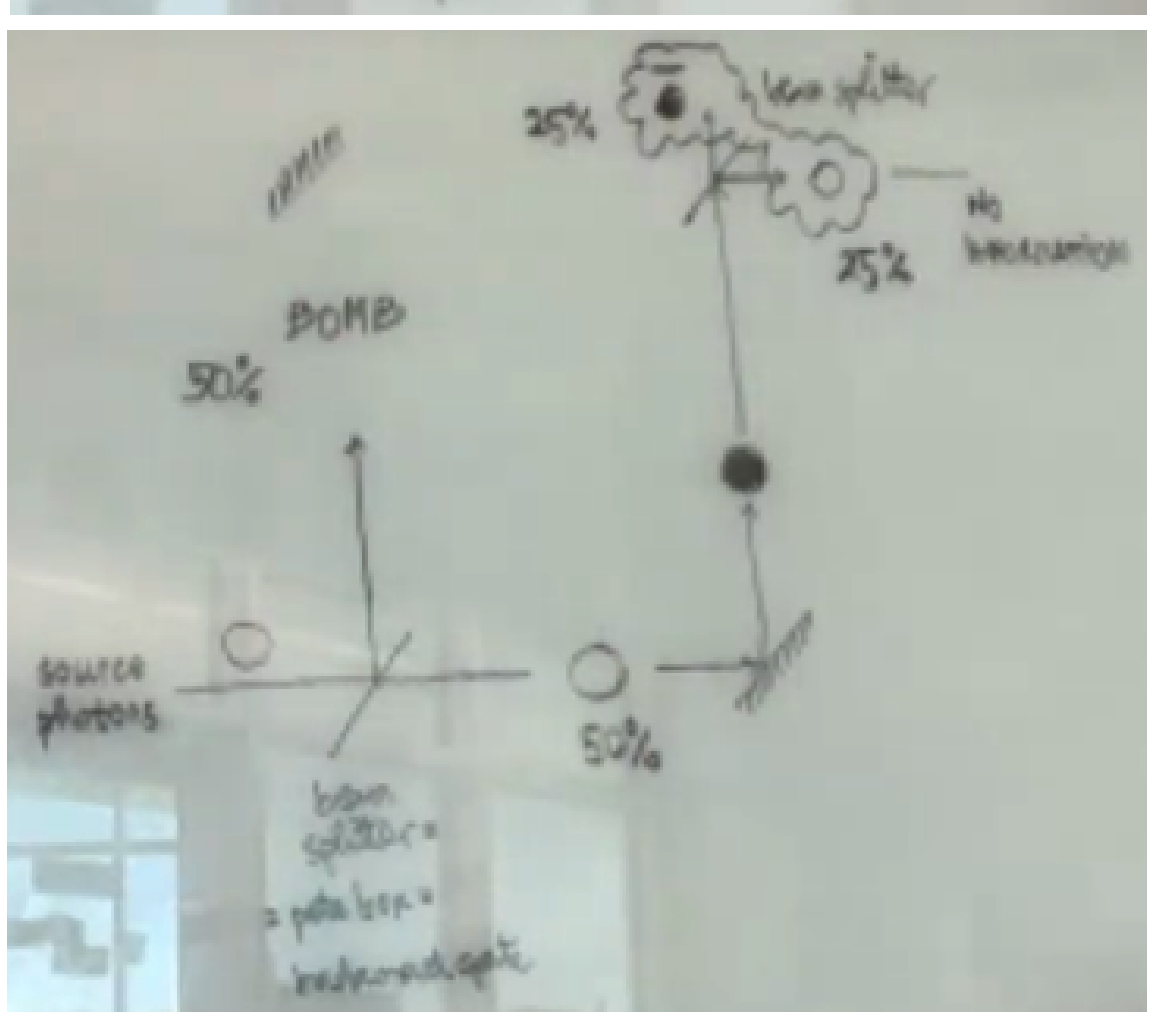

Figure 3: Misty states can explain the Flytrap experiment.

## 5 IBM Qiskit

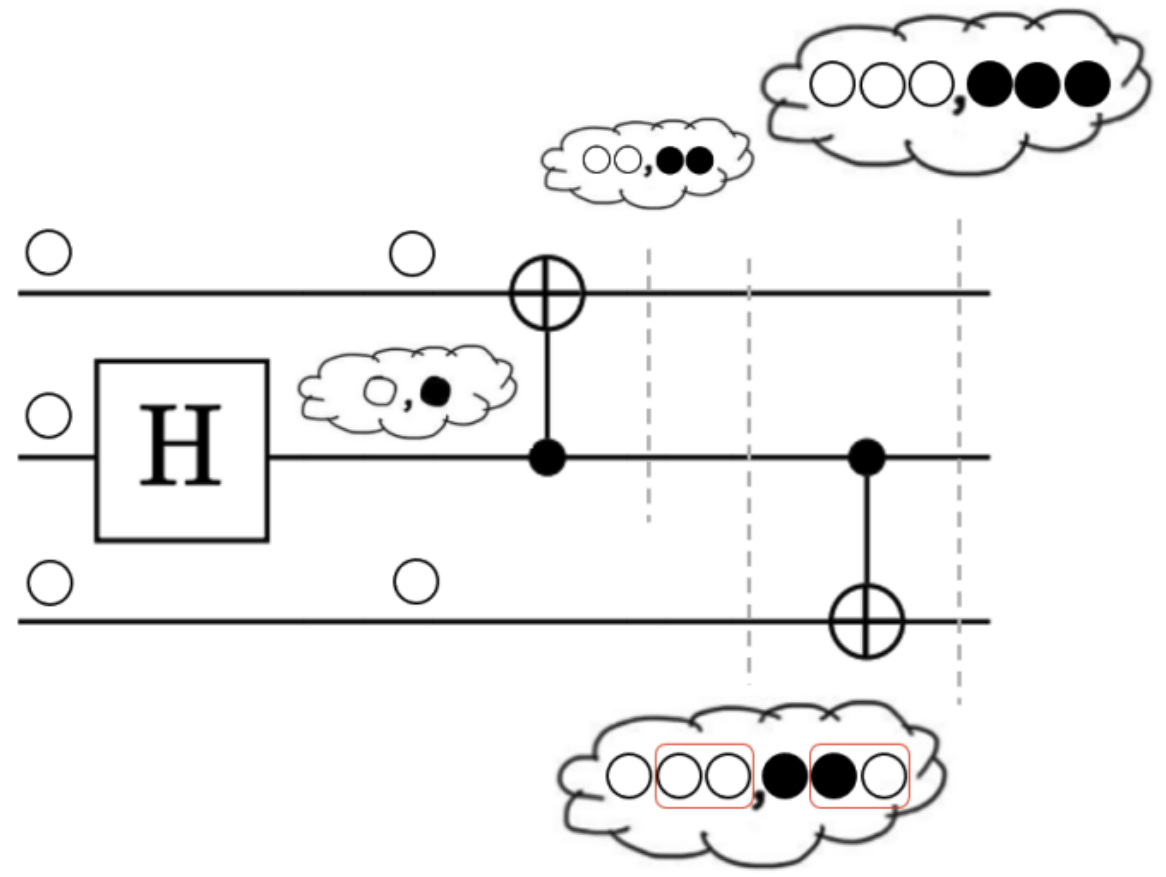

Figure 4: Misty states can explain the GHZ state circuit.

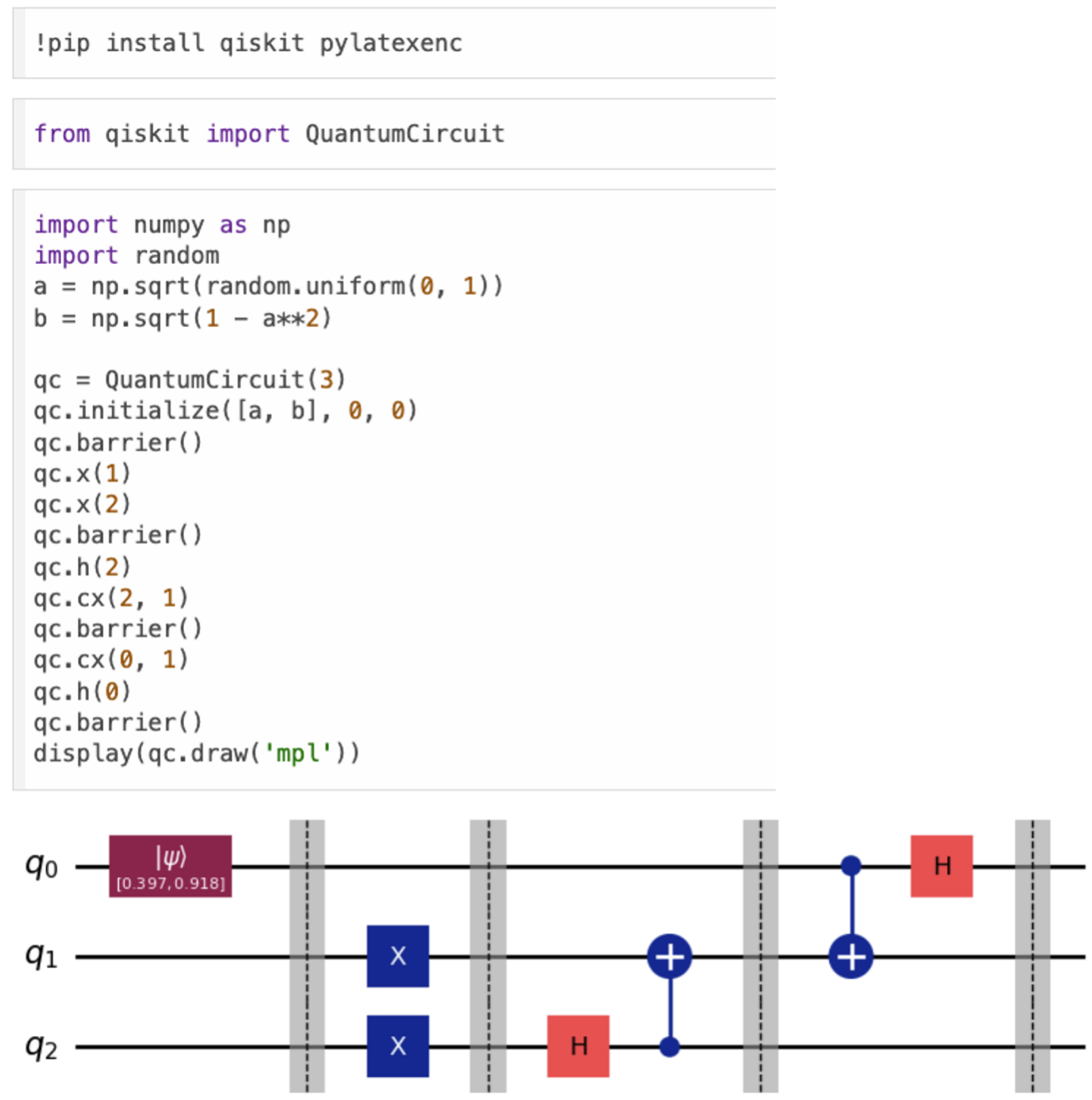

Figure 5: The Snake Equation as an IBM Qiskit circuit.

The Snake Equation is a fundamental equation in the ZX calculus. It says that (in the circuit above) if we measure the first two lines and obtain the same values as the values that are used as inputs in the last two lines then the third output line will carry the (arbitrary) state that is on the first input line of the circuit. This should immediately suggest quantum teleportation to the reader. Here's the experimental proof:

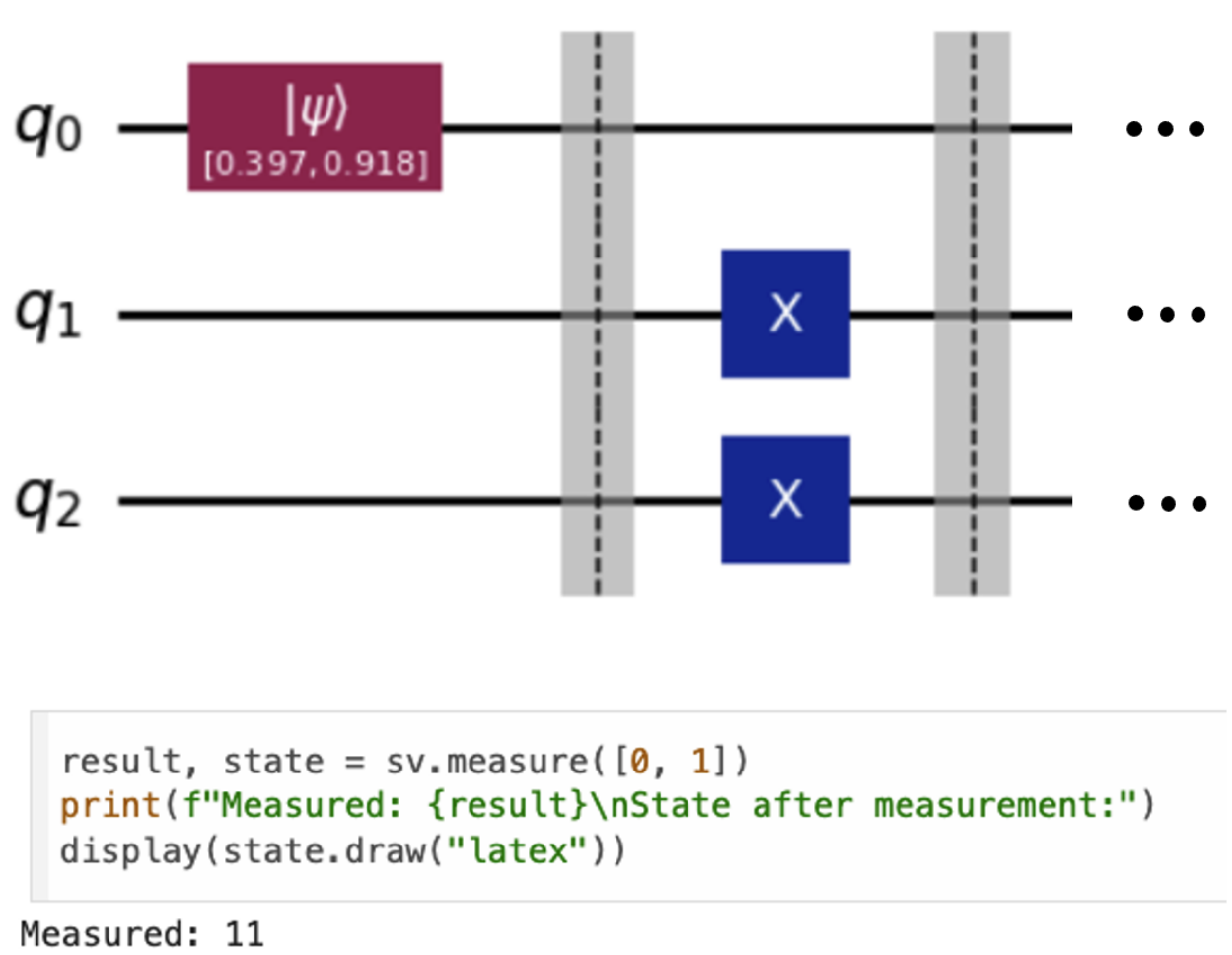

Figure 6: A close up of the experimental results.

## 6 Phase Kickback

We now refer to a well known relationship that has been proved by many authors using misty states. One such reference is [9] and here's the basic setup:

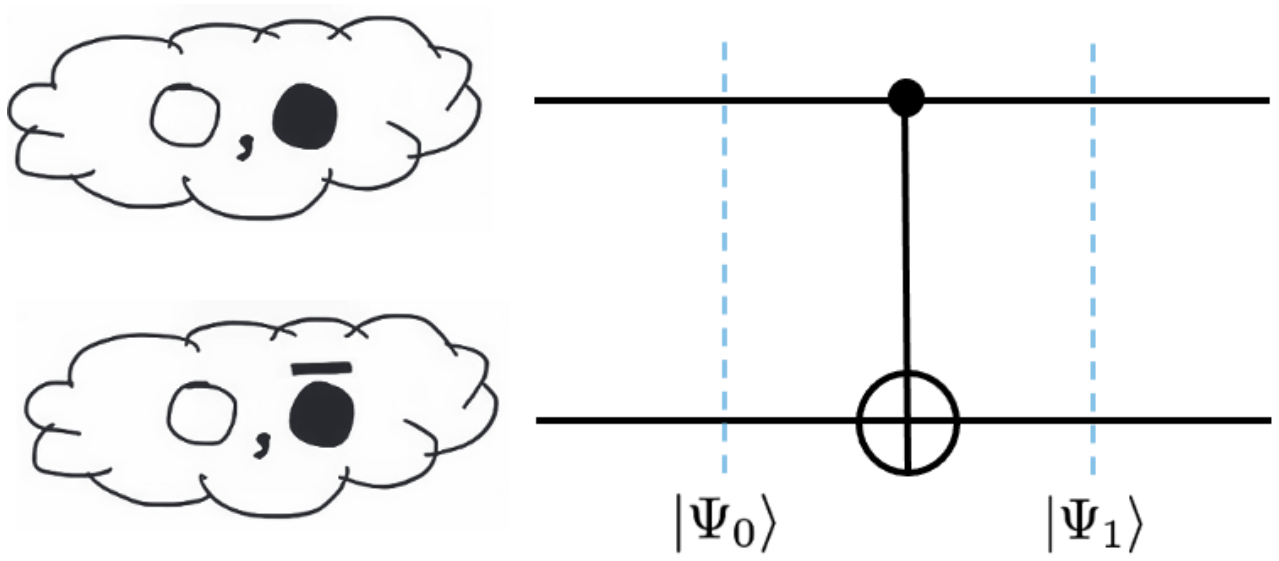

**Figure 7: When the control and the target reverse roles.**

Here's what happens in the figure above:

$$|\Psi_0\rangle = \{\{\bigcirc, \bullet\}, \{\bigcirc, \overline{\bullet}\}\} = \{\bigcirc\{\bigcirc, \overline{\bullet}\}, \bullet\{\bigcirc, \overline{\bullet}\}\}$$

We then have:

$$\begin{aligned} |\Psi_1\rangle &= \{\bigcirc\{\bigcirc, \overline{\bullet}\}, \bullet\{\bullet, \overline{\bigcirc}\}\} \\ &= \{\bigcirc\{\bigcirc, \overline{\bullet}\}, \overline{\bullet}\{\bigcirc, \overline{\bullet}\}\} \\ &= \{\bigcirc, \overline{\bullet}\}\{\bigcirc, \overline{\bullet}\} \end{aligned}$$

## 7 Quantum in Pictures

In this section we rely on ZX diagrams:

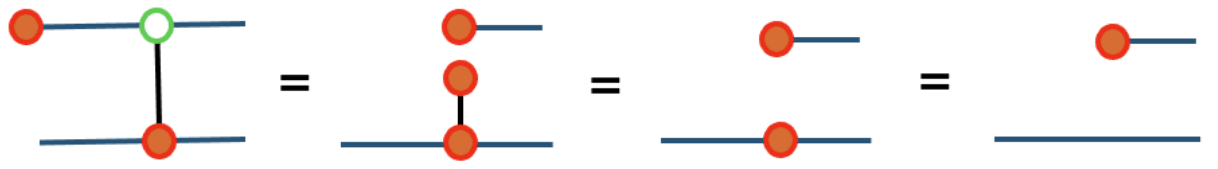

**Figure 8: A C-NOT gate when $|0\rangle$ is applied on the control**

The diagram above and the one below show the functionality of a C-NOT gate in the computational basis.

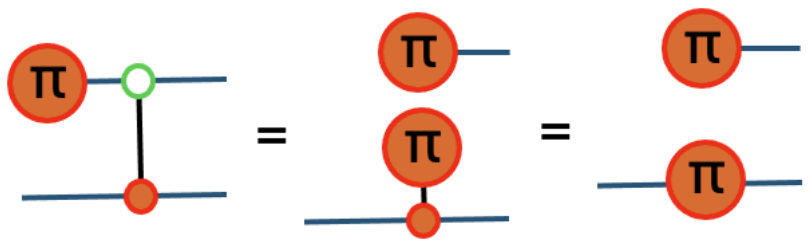

**Figure 9: With $|1\rangle$ on the control the target becomes a NOT**

The neat thing about the ZX calculus is that its scope is wider than that of the tools we have referred to thus far. Many dualities suddenly become accessible. If we change the basis (that is from $|0\rangle$ and $|1\rangle$ to $|+\rangle$ and $|-\rangle$, that is, if we swap the colors of the spiders) we obtain the same behavior just with the roles of the control and the target reversed. Phase kickback then becomes immediate and we get a quicker and more self-contained resolution to the Bernstein-Vazirani challenge than (as a comparison) what is shown in [9]. The details of this argument are presented in this section.

So here's first the derivation for phase kickback followed by the resolution of a Bernstein-Vazirani challenge. As a reminder, a green spider state with phase 0 is $|+\rangle$ and a green spider state with a phase of $\pi$ is $|-\rangle$. We then apply the copy rule, once, and fusion twice.

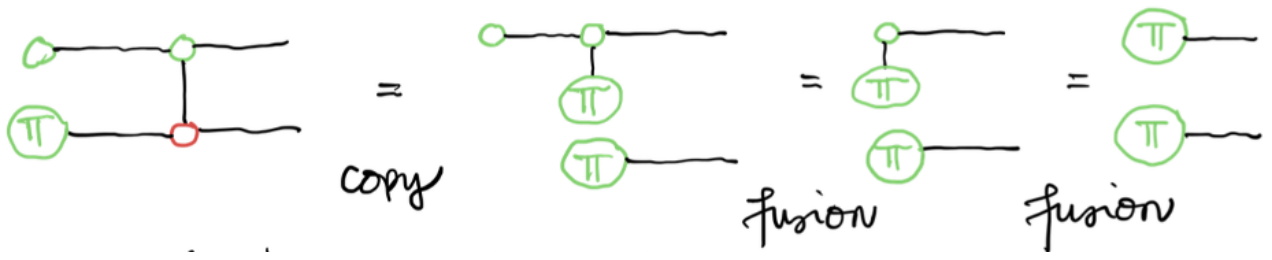

**Figure 10: The shortest possible proof for phase kickback.**

It should be clear then (or worth reminding) that the graphical rules in ZX are actually mathematical subroutines.

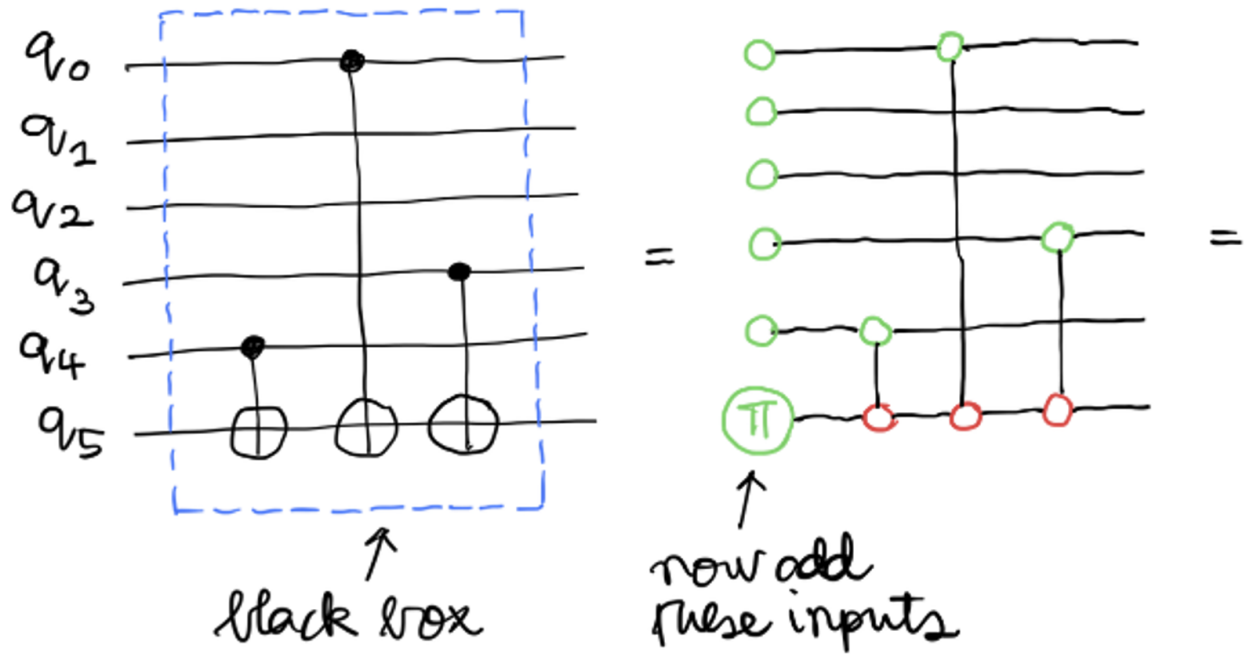

**Figure 11: Starting point for a Bernstein-Vazirani challenge.**

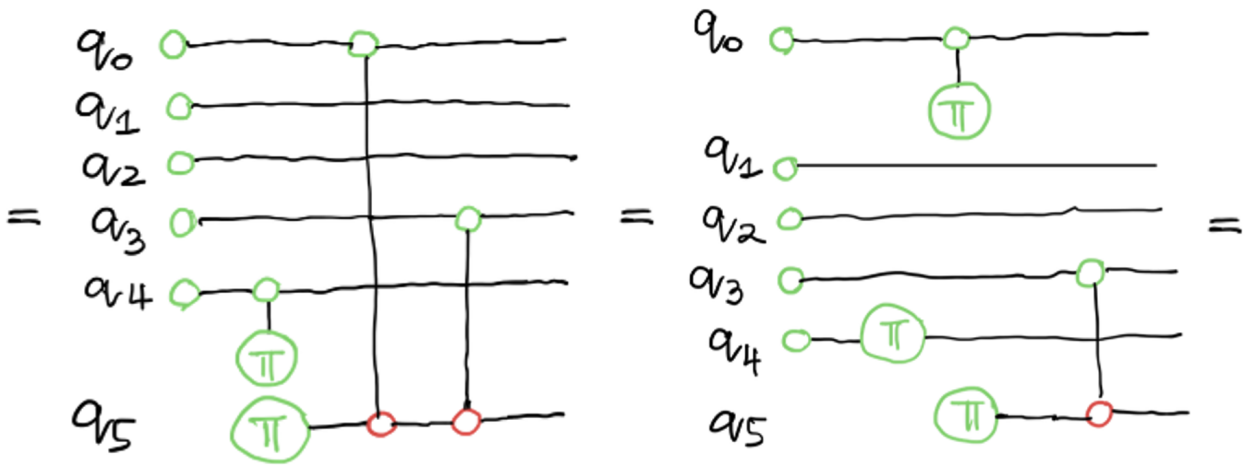

**Figure 12: We now show what happens inside the black box.**

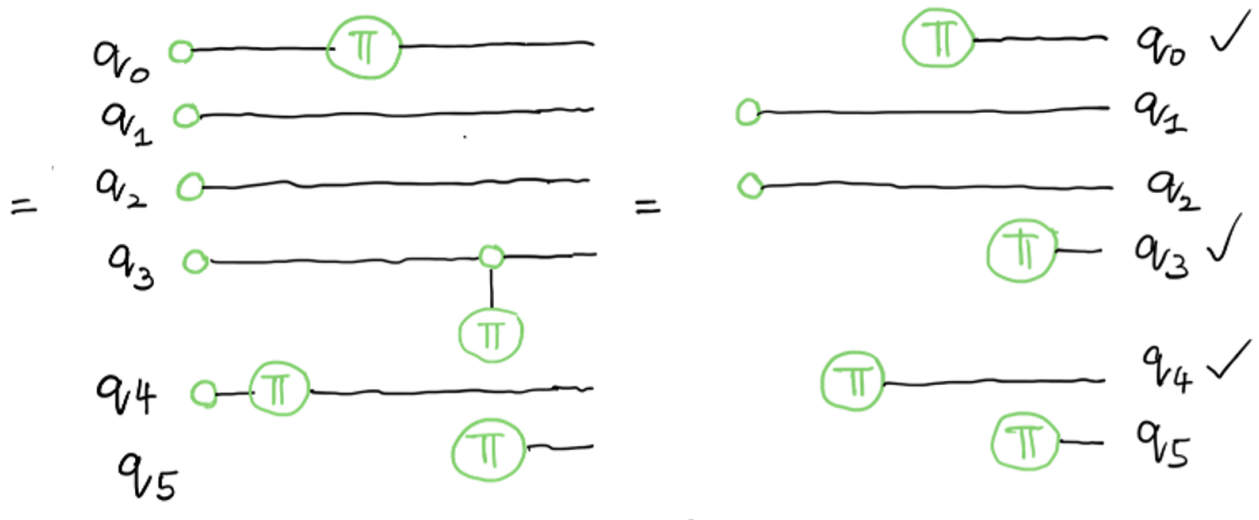

**Figure 13: And the process concludes with the answer(s).**

Thus, the black box reveals itself in the outputs in only one step.

## 8 Student Feedback

In this section we would like to mention the following alternative solution to Elitzur-Vaidman with polarizing beam splitters:

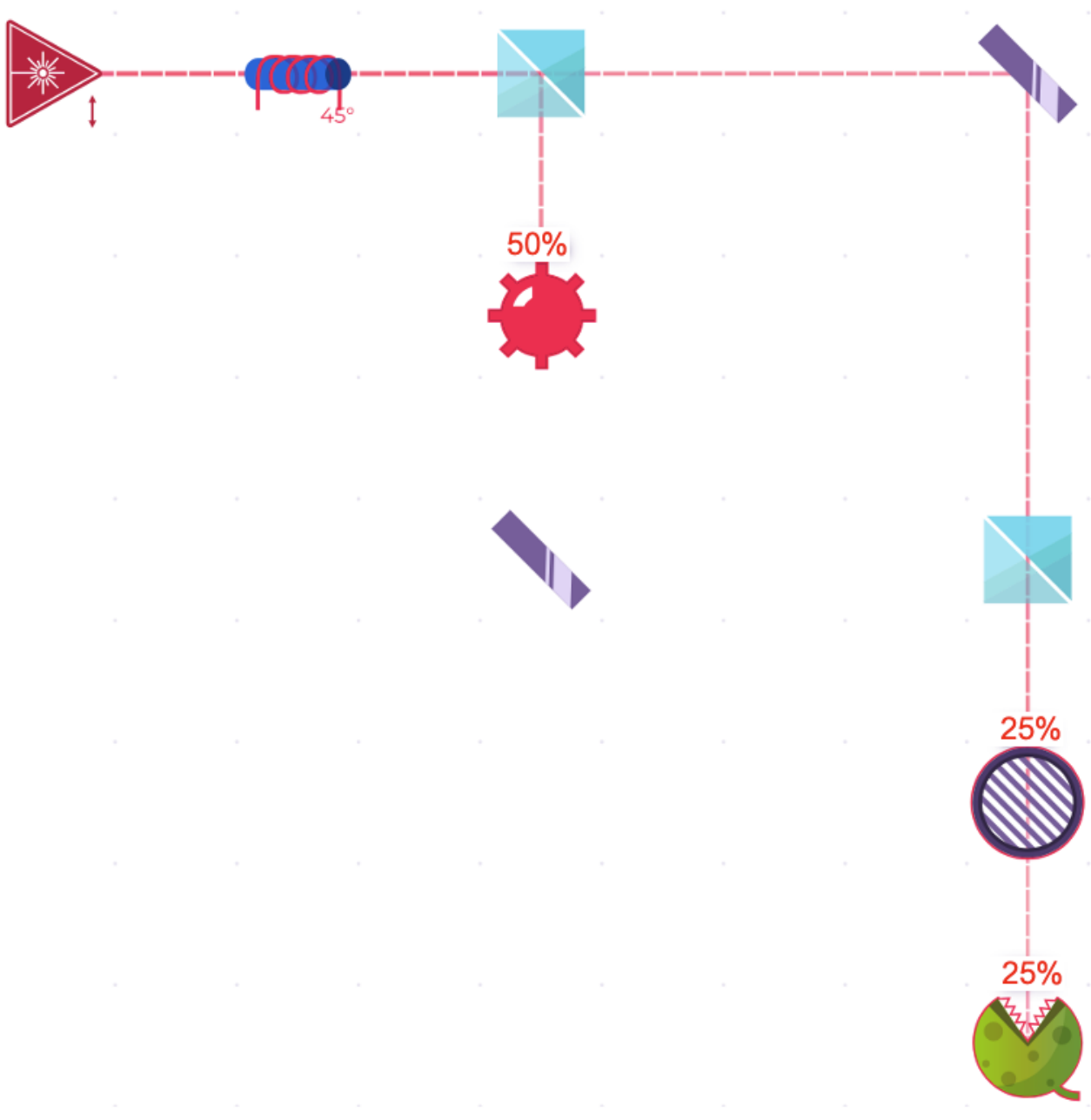

**Figure 14: Quantum seeing in the dark (student solution).**

Here's the reference setup:

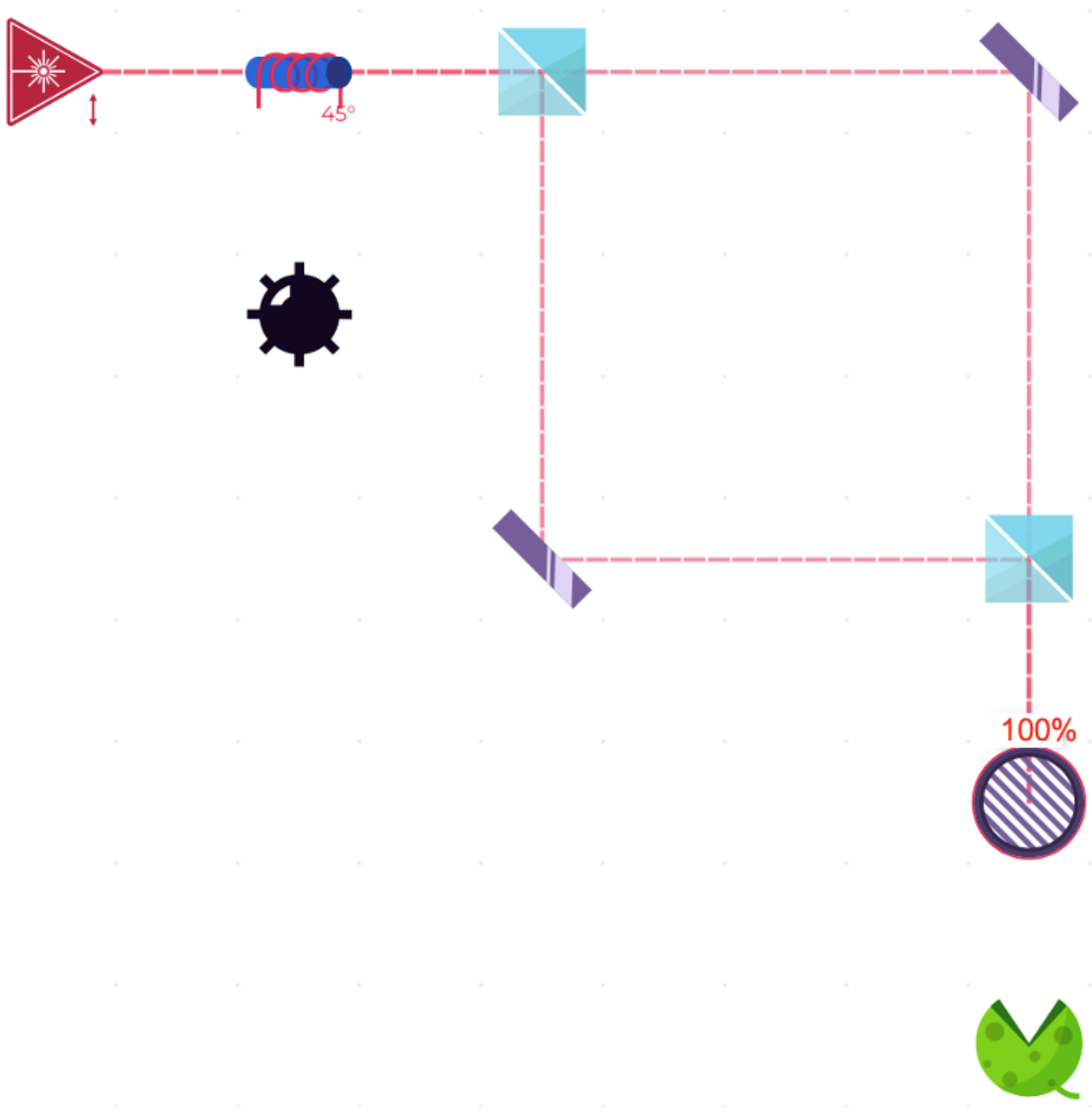

**Figure 15: Without a bomb the detector is always silent**

It's worth mentioning that the solution above was the result of interaction between a team of students none of whom was a major of Physics. In this regard it's also relevant pointing out that everybody enjoyed working with the Quantum Flytrap (students, whether in HS, undergraduate or graduate, including the HS teachers). The resource at [13] is at the same time very engaging, entertaining and also very effective. Students developed posters, made presentations (at local, school and state, but also national and international events), eventually leading into the Quantum Zeno effect (where probability of detecting the bomb without exploding it can be brought arbitrarily close to 100% (i.e., the certain event)).

The multiple formalisms and tools used represent a strength: starting with misty states and the quantum flytrap allows us to introduce the math (and some physics) without students being explictly aware of that. We then develop circuits in the emulator (IBM Qiskit) and reason about them using misty states. Once we reach the Qiskit circuit that represents the snake equation we start introducing the ZX rules and conventions. We continue to present topics but we look at everything from more than one perspective. One student remarked: "I really like that I have 3 or 4 ways to do quantum circuits and understand them, the linear algebra, the [D]irac notation, misty states which is easier to write than [D]irac for me and now [ZX] calculus which so far seems to have the most potential to be intuitive, at least for unitaries/circuits."

On a scale from 0 ("learned nothing") to 10 ("learned lots about quantum") HS students rated their quantum flytrap experience with a 7.4 and mentioned that sometimes the gameplay would take over and become addictive. By contrast their experience with the misty states formalism (supported by [16] and Python implementation of circuits in IBM Qiskit) was an 8.6 (so considerably higher).

Here are some more (from a selection of) answers collected earlier this summer (student responses appear in square brackets):

(1) What I liked most about this class was [[that i]t was a good chance to learn about something that I am int[e]rested in but that rarely gets taught to high schoolers].
(2) What I liked least about this class was [[that] I wish we had more time to code because it felt sort of incomplete.]
(3) The thing I am most proud of having learned in this class was [[h]ow to code with quantum ideas.]
(4) Before this class I thought quantum physics was [way more complex to even think about,] but now I think [you just need to be curious and have an open mind to conceptualize it].
(5) Something I learned about quantum gates is: [[t]hat you can a have certainty in an uncertain quantum problem.]
(6) Shor's Algorithm—what did you get from the video and discussion? [I now know that encryption as we know it today is not going to hold up very long and it will be cool to see where it goes from here. I'm not even lying when I say this, I'm writing my government paper over how quantum computing will change the way we encrypt our messages and what's been tried to change the way we encrypt our messages. So thanks for an essay idea haha.]

Answers collected were encouraging even in those cases where the message was highly nuanced. As an example a student wrote: Before this class I had heard that quantum physics was [awesome, but confusing.] Now I think [it] is [confusing, but awesome].

## 9 Learning Outcomes

The kind of learning outcomes we're aiming for here are slightly more general than usual as we'd like these dual credit classes to apply both at the HS level as well as to incoming college students[1]. So, as a general rule, we'd like these classes to provide General Education (GenEd) credit as well as, more specifically, either Natural Science and Mathematics (N&M) credit or Mathematical Modeling credit. Courses in N&M usually expose students to the nature and methods of scientific inquiry, emphasizing quantitative approaches to the testing of falsifiable hypotheses. These courses begin to provide students with the tools and skills required not only to understand physical and biological phenomena but also to discover them through theoretically based inquiry, rigorous analytical thinking, and/or the collection and interpretation of empirical data, broadly interpreted. Development of these skills is essential for preparing students to be informed and active participants in modern society.

Students who complete the N&M requirement will demonstrate: (a) an understanding of scientific inquiry and the bases for technology; (b) the ability to model and understand the physical and natural world; (c) the ability to collect and interpret data, think critically, and conduct theoretically based inquiry; (d) the ability to solve problems; and (e) analytical and/or quantitative skills.

Mathematical Modeling courses provide rigorous instruction in fundamental mathematical concepts and skills presented in the context of real-world applications. The modeling skills provide analytical methods for approaching problems students encounter in their future endeavors. What is very important here is the significant overlap between the mathematics needed for quantum computing and the mathematics needed for AI (Learning) and Data Science. The ultimate goal of the dual-credit classes is to make students aware of, and introduce them to complex numbers and linear algebra (along with the postulates of quantum mechanics). In this sense [12] is a meaningful target, being the support used at the Quantum School for Young Students at the University of Waterloo.

## 10 List of Proposed Topics

The following is a chronological list of proposed topics:

- What is a Qubit? The misty states formalism. Superposition, interference, phase. Quantum gates.
- Quantum Game. Quantum Flytrap. The Elitzur-Vaidman experiment. The Quantum Zeno effect.
- Two-qubit quantum gates. Quantum circuits. What is entanglement? Introduction to IBM Qiskit.
- Case study: phase kickback and Bernstein-Vazirani.
- Case study: Deutsch-Josza.
- Case study: Grover search algorithm.
- Case study: Superdense coding.
- Case study: Entanglement swapping.
- Case study: Quantum teleportation.
- Introduction to the ZX Calculus (Quantum in Pictures)
- What is the CHSH game?
- Case study: the GHZ game. Non-locality.
- Quantum error correction (bit-flip, phase-flip, Shor code).
- Cybersecurity alert: the Shor algorithm.
- Post quantum encryption. The NIST framework.
- Quantum Key Distribution (QKD).
- Quantum Bayesianism (QBism)
- The mathematics of quantum computing

We have taught this class for 5 years now and believe this selection of topics to be both feasible and appropriate for the purposes of our class. Note also that this one college class will likely result in three HS classes carrying the same amount of total credit.

[1] Ideally, these students would take these classes while in HS and then come with advanced credit straight into our quantum computing specialization.

## 11 Next Steps and Future Work

In our experience this project was both worth pursuing and is now close to being finalized successfully. As a side effect we aim to add Quantum to the HS curriculum for our state. We then aim to create a task force of interested stakeholders at the national level (from the industry, academia, national labs and other government agencies) and help College Board develop materials for a course, curriculum and AP exam for Quantum Information Science (QIS).

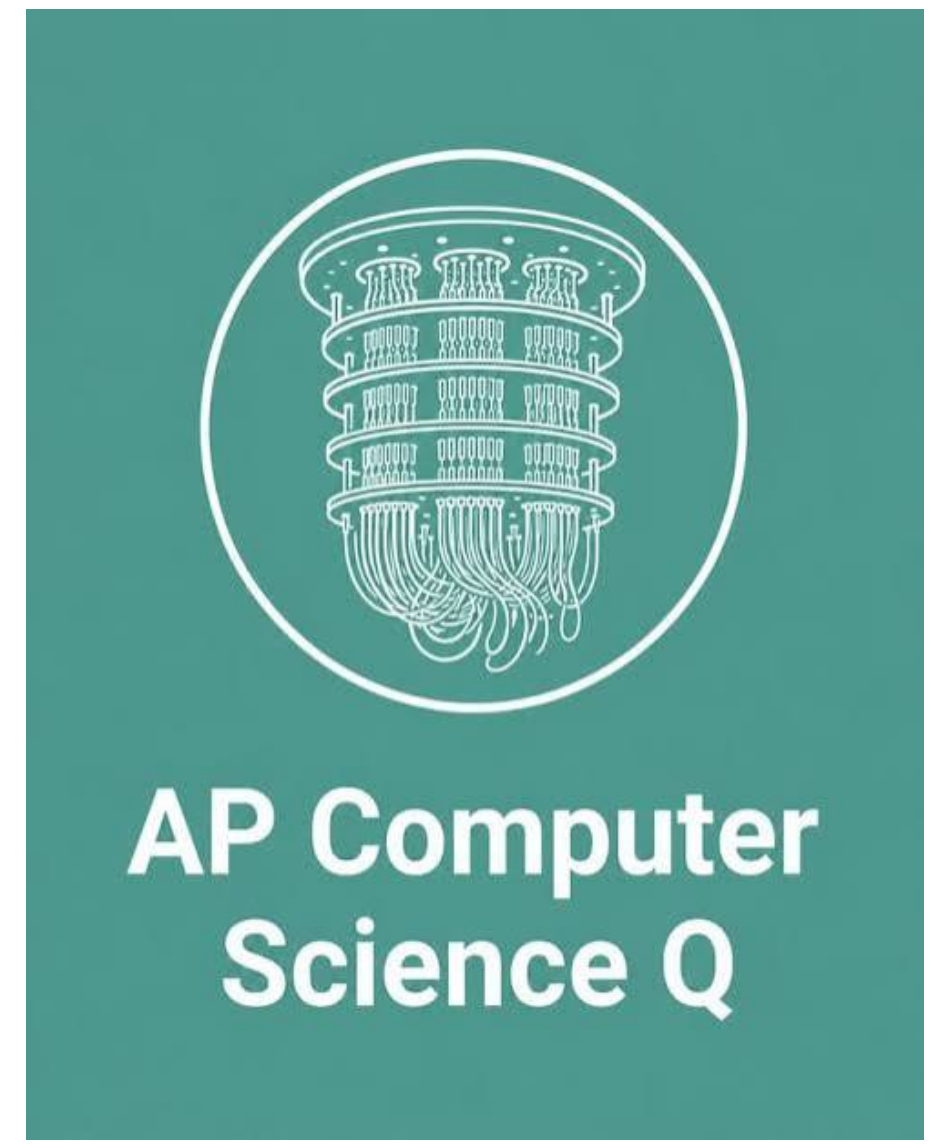

**Figure 16: Inspiring artwork (for us) by unknown author**